\documentclass{bmvc2k}

\usepackage{amssymb}
\usepackage{graphicx}
\usepackage{subfigure}
\usepackage{url}
\usepackage{epstopdf}
\usepackage{amssymb}
\usepackage{booktabs}
\usepackage{mathtools}
\usepackage{enumerate}
\usepackage{CJK}
\usepackage{url}
\usepackage{algorithm}
\usepackage{algorithmic}
\usepackage{multirow}
\usepackage{nomencl}
\makenomenclature
\usepackage{color}

\urldef{\mailsa}\path|huibo.bi12@imperial.ac.uk|

\title{Emergency and Normal Navigation in Confined Spaces}

\addauthor{Student: Huibo Bi}{huibo.bi12@imperial.ac.uk}{1}
\addauthor{Supervisor: Prof. Erol Gelenbe}{e.gelenbe@imperial.ac.uk}{1}

\addinstitution{
 Intelligent Systems and Networks Group\\
 Department of Electrical and Electronic Engineering\\
 Imperial College London\\
}

\runninghead{Huibo Bi}{Emergency and Normal Navigation in
Confined Spaces}

\begin{document}

\maketitle

\section{Introduction}
My research interests lie in the field of emergency navigation and emergency management of confined places in both standard and emergency conditions. In conjunction, novel routing algorithms to realise energy efficiency in evacuation process are investigated.

Disasters such as fire in buildings or vessels could cause a large amount of injuries and fatalities as well as enormous losses in properties. This problem aggravates in recent decades due to the urbanization in modern cities. To mitigate the loss during a disaster, fire drills are held to facilitate evacuees gain experience from simulated scenarios. However, the effectiveness of the fire drills is low due to drawbacks such as time-consuming and they can not model the actual situation in a disaster \cite{filippoupolitis2010emergency}.

With the rapid development of the relevant technologies and the advent of low-cost wireless sensor networks, emergency response systems have become a significant component of large scale physical infrastructures \cite{gelenbe2012large}. Therefore, related navigation algorithms, asset allocation algorithms, reliable secure communication approaches \cite{gorbil2012resilience}, models and tools for emergency prediction and simulation have motivated considerable research. As the core of the emergency response systems, navigation algorithms aim to evacuate civilians to the exits safely and efficiently. However, due to the highly dynamic nature of an emergency environment, current state-of-the-art navigation algorithms for emergency still face a lot of challenges. For instance, most algorithms need a new convergence procedure or a full graph search when the hazard spreads. This process can be time-consuming due to the transmission interference and network congestion in communication \cite{li2003distributed}. Meanwhile, the communication overhead will increase exponentially with the scale of the building layout. Another problem is that most algorithms tend to find either the shortest paths or the safest paths but do not consider the impact of congestion, which is a vital factor in high occupancy rates \cite{bib:BiDesmetGelenbeISCIS2013}. Current congestion-aware algorithms can be divided into congestion-free algorithms and congestion-ease algorithms. Congestion-free algorithms which are based on the minimum cost network flow problem \cite{ford2010flows,ahuja1988network} are impractical due to the high computational overhead and assumption of a non-spreading hazard. Congestion-ease algorithms employ queueing networks models \cite{lino2011tuning,desmet2013graph}, flow-based models \cite{wang2008modeling,pizzileo2010new} or potential-maintenance models \cite{chen2008load,chen2011distributed} to estimate the time delay on a path and redirect evacuees accordingly. However, the effect of current congestion located on a path is not taken into account. Furthermore, most evacuation routing algorithms are not very fault-tolerant to the garbled messages as the decisions are made completely on newly obtained data and little experience is gained from previous decisions.

The convergence process hinders the performance of a distributed system by information synchronization \cite{gelenbe1979analysis} and a full graph search is not necessary because the decisions of an emergency navigation algorithm mainly depend on information provided by strategically located sensors \cite{desmet2013interoperating}. Therefore, a ``self-adaptive'' routing algorithm originally used in large-scale and fast-changing packet networks \cite{bib:BiDesmetGelenbeISCIS2013} is employed in the sequel to optimise the evacuation process. The approach namely ``Cognitive Packet Network'' (CPN) can achieve fast adjustments because each node operates as a sub-system and no synchronization procedure is required. Smart packets are used to explore paths with both its own observations and experience of predecessor packets. Hence, routes can be refined gradually and communication overhead is reduced markedly. Meanwhile, computational load can be decreased significantly as the calculation only focuses on the path collection discovered by SPs.

Furthermore, previous emergency navigation algorithms concentrate on ``normal'' evacuees and use a single metric such as path length or safety level to guide civilians. For instance, most flow-based models hypothesise evacuees as continuous flows and all civilians are homogeneous. The ``Magnetic model'' in \cite{okazaki1993study} sets a walking velocity for each evacuee but ignores personal requirements as well as the effect of social interaction. Although the advent of multi-agents models makes it more convenient to customise physical attributes for individuals, most research focus on incorporating sociological factors due to the easiness of describing social behaviours such as coordination or stampede in multi-agents models. In fact, there are commonly diverse categories of evacuees such as adults, children, people in wheelchairs and ill people in an evacuation process. Due to the difference in mobility and resistance to hazard, they should be routed with regard to different Quality of Service (QoS) needs. In contrary to the previous algorithms, CPN based algorithm can suffice several kinds of evacuees' specific QoS criterion because each SP can be sent out with a unique predefined QoS goal and brought back routes with respect to a distinct metric.

In the following sections, a literature review on emergency navigation algorithms and related technologies to realise energy efficiency in emergency are presented in section 2. Current work on CPN based algorithm with diverse metrics and the corresponding experimental results are reported in section 3. Finally, the future research plan is given in section 4.

\section{Literature Survey}

Originating from the egress problem \cite{weinroth1989adaptable}, emergency navigation is the process of guiding evacuees to exit when a disaster happens. This section will present an overview of current research on emergency navigation algorithms. Additionally, corresponding approaches to realise energy efficiency in robots, mobile sensor networks as well as wired or wireless packet networks that have some reference significance in realising energy efficiency emergency evacuation are also reviewed.

Emergency evacuation of a confined place is a complicated and transient transshipment problem due to the spreading of the ongoing hazard, errors and delays in communication, movement and congestion of evacuees, malfunction of communication infrastructures, etc. Therefore, it is a great challenge to find and allocate appropriate paths in a real time fashion. Currently, there are six main approaches to compute paths: potential-maintenance approaches; geometric or graph theory based approaches; prediction-based approaches; network flow optimisation approaches; biological-inspired approaches and adaptive QoS aware routing inspired approach \cite{gelenbe2012large}.

Ref. \cite{li2003distributed} presents a self-organizing sensor network to guide a user (robots, people unmanned aerial vehicle, etc.) through the safest path by using ``artificial potential fields''. The dangerous zones are considered as obstacles. An attractive force is generated by the destination to pull the user while a repulsive force is generated by the obstacles to push the user away from them. Experiments are conducted on a 50 Mote MOT300 testbed and the results indicate that the algorithm could successfully direct the objects to the destination. However, multiple destinations may affect the efficiency of reaching the exits as the users move under the actuation of artificial forces. The convergence time for network stabilization is relatively long due to the effect of data loss, asymmetric connection and network congestion. Meanwhile, this algorithm has not been tested on a large scale area. Ref. \cite{tseng2006wireless} proposes a temporally ordered routing algorithm \cite{park1997highly} based multipath routing protocol to route evacuees to exits through safer paths. A navigation map is manually defined to avoid impractical paths and each sensor is assigned with an altitude with respect to its hops to the nearest exit. When an emergency event is detected, the sensor sets its altitude to a high value and routes are formed from higher altitudes to lower altitudes. In addition, hazardous regions are constructed by sensors within a predefined hop distance from a hazard. Ref. \cite{pan2006emergency} extends the algorithm in Ref. \cite{tseng2006wireless} to a 3D environment and divides the sensors to several types in terms of location. However, communication requirement of this class of approaches is difficult to fulfil in a hazard and the ``local minimum'' problem need to be considered.

The localized Delaunay Triangulation method \cite{li2002distributed,li2003localized} is adopted in \cite{chen2008distributed} to partition a wireless sensor network into a triangular area and construct a mount of area-area paths. The level of the hazard in a triangle is classified by the readings of sensors inside it. A load-dispersion algorithm is employed to distribute evacuees by limiting the number of users per exit. Without location information of sensors and users, Ref. \cite{li2009sensor} uses a load map constructed by a set of points which are closest to at least two sensors on the margins of hazardous area to form backbones. Evacuees around the dangerous areas are directed to the backbones with a virtual power field assigned by measuring signal strength. Combining the definition of effective length with Dijkstra's algorithm, Ref. \cite{filippoupolitis2009distributed} builds a decentralized evacuation system which comprises of decision nodes (DNs) and sensor nodes (SNs) to compute shortest routes in real time. Ref. \cite{gorbil2012intelligent,gorbil2013resilient} propose an opportunistic communications \cite{pelusi2006opportunistic} based emergency evacuation system which includes SNs and mobile communication nodes (CNs). Dijkstra's shortest path algorithm is used by CNs to calculate the shortest path to exits. Ref. \cite{lu2003evacuation} employs a variation of Dijkstra's Shortest Path algorithm in which the edge travel time is used as metric. The node capacity and edge capacity are considered to avoid congestion. The current available edge and node capacity are recorded as time series and vary with reservation. A heuristic algorithm namely Single-Route Capacity Constrained Planner runs continuously to determine the number of evacuees in a flow with regard to the minimum node or edge capacity on the path. It also reverses capacity for the flow according to the expected arrival time and departure time. This algorithm can reduce stampedes significantly at cost of very high computational overhead. Meanwhile, precise estimation of arrival time and departure time is difficult in a complex environment.

By adopting a predetermined ``Detector Response Mode'' \cite{olenick2003updated} to predict the spread of hazardous conditions and the speed of evacuees, Ref. \cite{barnes2007emergency} maintains a ``hazard graph'' and a ``navigation graph'' to compute the summation of minimum intervals between the arrival time of the fire and an evacuee at each node on a path. The path with the minimum time interval is defined as the safest path. Ref. \cite{berry2005firegrid} presents a Monte-Carlo-based fire model to forecast an uncontrolled compartment fires. In conjunction with real time sensor data, a super-real-time simulation is developed to estimate the spread of fire and smoke throughout a building.

By considering the evacuation problem as a minimum cost network flow problem \cite{ford2010flows,ahuja1988network}, EVACNET+ \cite{chalmet1982network,francis1984negative,kisko1985evacnet+} converts the evacuation network to a time-expended network in which original network is copied in time units. The time-expended network is solved to get the optimal solution. Polynomial time algorithm is employed in Ref. \cite{hoppe1994polynomial,hoppe2000quickest} to find the quickest flows between a fixed number of sources and sinks. The minimum cost flow is resolved by ellipsoid method in a finite number of steps. However, the computational overhead of those algorithm is extremely high and is impractical for real applications.

A feed-forward neural network model is adopted to a wireless sensor-actuator network (WSAN) for evacuation routing in Ref. \cite{jankowska2009wireless}. All physical nodes in the WSAN deploy a neural network with identical topology: an input layer, a hidden layer and an output layer. The input layer receives the latest two coordinates of a pedestrian and a suggested direction is subsequently generated by the output layer. The neural networks are trained with a back-propagation algorithm \cite{mitchell1997machine} in standard situations and are deactivated when a emergency happens. Hence evacuees will be directed to exits over their normal walking paths. However, back-propagation algorithms suffer from slow learning rate and easily converging to local minimum. Furthermore, this model cannot react to the spread of a hazard. Ref. \cite{li2010multiobjective} employs a genetic algorithm \cite{john1992holland} to minimise total evacuation time, travel distance and number of congestion. Non-domination sorting \cite{deb2000fast} is used as no priori knowledge is available to determine the weight of the three goals. The initial chromosomes are paths found by the k-th shortest path algorithm \cite{eppstein1998finding} and incrementally evolved to feasible solutions through crossover and mutation. As a evolutionary approach, this algorithm has advantages in solving multi-objective optimization problem (MOP) \cite{saadatseresht2009evacuation}. However, the computational overhead is high due to the path-finding and the evolution process. Ref. \cite{pan2005multi} adopts a variation of particle swarm optimization (PSO) to search routes and adjust velocity during evacuation. Occupants are viewed as particles to search exits. Once an exit is found, all the other particles will move towards it while keep their moving inertia to expand searching space. If more than one exit is found, particles will choose the nearest exit as the moving target. Nevertheless, use occupants directly to explore paths may cause high fatality rate. Meanwhile, this algorithm may suffer from seriously congestion and oscillation problem. Inspired by the bee colony foraging behaviour, Ref. \cite{samadzadeganbiologically} uses bee colony optimization \cite{karaboga2005idea} to displace evacuees in hazardous areas to safe areas during an emergency. Hives, food sources and bees represent safe areas, hazardous areas and evacuees respectively. Evacuees select a safe area according to ``attractiveness'' which is determined by the distance to the area and the number of people in a hazardous area. Once a evacuee determines a target, it will recruit other evacuees by sharing information of the devoted area. This algorithm obtains a robust evacuation plan at the expense of high communication overhead.

Ref. \cite{filippoupolitis2010emergency} presents a decision support algorithm inspired by CPN routing protocol \cite{gelenbe2001towards} \cite{gelenbe2001design}. SPs are sent to find the routes with potential best QoS and acknowledgement packets (ACKs) are used to return information gathered by SPs. SPs' next hop is decided by m-Sensible routing algorithm \cite{gelenbe2003sensible}. Rolling average mechanism is used to obtain the new effective length from the previous value and current reported one. In order to maintain network stabilization, a predefined ``measurement discard threshold'' is set to discard the reported effective lengths with only small change. One shortcoming of this algorithm is that little experience is cumulated from the continuously routing procedure and a new path selection is independent from the previous one.

Achieving energy efficiency is a significant goal in emergency evacuation because depletion of power for specific evacuees such as disabled people in wheelchairs can be fatal. Meanwhile, disasters will affect infrastructures and cause difficulties for evacuees. For instance, a fire will typically cut off the lights and slow down evacuation procedure and cause congestion.

To optimise energy utilization for specific evacuees such as the disabled in wheelchairs, related research on robotics and mobile sensor networks can be used as references. An energy-efficient motion planning approach is presented in \cite{mei2004energy} by modelling the relationship of the motor's speed and the power consumption with polynomials. Acceleration and turns are considered in this approach. Optimal paths on terrains are computed for a mobile robot in \cite{sun2003energy} to minimise energy consumption and avoid impermissible directions with overturn danger or power limitations. Energy is assumed to be expended by both friction and gravity. Different velocity schedules for minimizing energy consumption is considered in \cite{wang2005optimizing} due to variable road conditions. Since in mobile sensor networks, mobile nodes tend to stop quite frequently, the effect of acceleration is taken into account. A weight $\frac{coverage \times distance}{power}$ determined by the degree of coverage, the distance to the target and the remaining power of a mobile sensor node is presented in \cite{verma2006selection} to decide which sensor to explore the location. Ref. \cite{wang2007exploring} proposes a centralized sensor dispatching algorithm (CentralSD) to dispatch mobile sensors to event locations while reducing the total energy cost and balancing the power utilization of each sensor. Maximum bipartite matching approach is used to allocate each sensor with a event location. To reduce message transmissions, the sensing field is divided into grids and CentralSD performs in each grid.

The enormous power utilization of Information and Communication Technologies (ICT) \cite{gelenbe2009reducing} has motivated considerable research in realising energy efficiency in wired or wireless packet networks. Because most of the emergency evacuation systems are based on a wireless sensor network and an evacuation process is similar to a network routing process, hence related research can provide insightful and innovative ideas for emergency evacuation. Ref. \cite{gelenbe2011framework} has given a comprehensive literature review of the state-of-the-art technologies on reducing energy consumption in this area. In particular, Ref. \cite{gelenbe2009reducing} presents a CPN based algorithm to minimise energy consumption while constrains the delay at an acceptable level. The goal function to be optimised by CPN is based on the relationship between power consumption of routers and the packet rate \cite{mahadevan2009power}.  Similarly, a CPN based power-aware routing algorithm is proposed in \cite{gelenbe2004power} to maximize the remaining battery life time while respecting the packet latencies by using a joint metric. Meanwhile, the CPN based algorithm can naturally decrease the energy consumption because smart packets unicast most of the time other than multicast.

Ref. \cite{gelenbe2011routing} and \cite{gelenbe2011framework} employ a G-networks model \cite{gelenbe1994g} with positive customers \cite{gelenbe1991product} and triggers \cite{gelenbe1993g} to imitate user packets and control packets respectively. Gradient descent optimization is used to optimise a energy utilization function by adapting re-routing probabilities of a certain category of user packets at a router.

\section{Current works}

My work to date has been primarily focused on adopting CPN, which has been widely applied in many areas\cite{sakellari2010cognitive}, to emergency navigation. In particular, diverse goal functions have been developed to fulfil different evacuees' QoS needs. The rest of the section is organized as follows: section 3.1 is a brief overview of my current work; a concept of CPN is introduced in section 3.2; the simulation model is described in 3.3 and the results are given in 3.4; the specific goal functions and corresponding experimental results are presented in 3.5.

\subsection{Overview}

Current state-of-the-art emergency navigation algorithms still face many challenges. For example, most algorithms need a full graph search or a convergence process, which induces high computational and communication overhead; most algorithms also assume that civilians have the same QoS need. On top of that, most algorithms are not fault-tolerant and can become unstable with the effect of incorrect reading of sensors. Furthermore, decisions are mainly depend on the current data and little experience is accumulated from the feedback of previous decisions.

To solve these problem, a CPN protocol is adopted based on recurrent Random Neural Networks \cite{gelenbe1993learning} to develop a decentralized, computationally efficient, fault-tolerant emergency navigation algorithm with respect to diverse categories evacuees' specific QoS needs. Random Neural Network (RNN) \cite{gelenbe1989random,gelenbe1990stability,gelenbe1991global,gelenbe1993learning} is a simplified model of biological neural networks. A random neural network model consists of neurons with a potential greater or equals to zero. If a neuron's potential is positive, it can fire a positive or negative spike at random interval to linked neurons or outside. A positive signal can increase the potential and negative signal can reduce the potential when it is positive. Random Neural Network has been successfully applied to a large amount of applications \cite{bakirciouglu2000survey} such as associative memory \cite{gelenbe1991associative,gelenbe1991distributed}, optimization \cite{gelenbe1993dynamical,gelenbe1992minimum,ghanwani1994qualitative}, texture generation \cite{atalay1992random,atalay1992parallel}, Magnetic resonance imaging \cite{gelenbe1996neural}, function approximation \cite{gelenbe1999function,gelenbe2004function}, mine detection \cite{gelenbe1997sensor,abdelbaki1999random}, automatic target recognition \cite{bakircioglu1998random,bakircioglu1998feature}, video compression \cite{cramer1996low,gelenbe1996traffic,cramer1998image}, network routing algorithms \cite{gelenbe1999cognitive,gelenbe2011self}, task assignment and emergency management \cite{gelenbe2010fast,gelenbe2010random,gelenbe2010randomemergency}.

The CPN based algorithm is implemented in a Java based simulation tool. When a simulation starts, certain number of agents based occupants with randomized initial locations begin to evacuate from a three-storey graph-based building scenario by using the CPN based algorithm. The performance of the algorithm is evaluated from the number of survivors, average health or other metrics. Other algorithms such as Dijkstra's algorithm are also executed for comparison purpose. Based on a congestion management algorithm, several different goal functions are developed to fulfil diverse categories of evacuees. Meanwhile, the oscillation problem in CPN is investigated and comparison experiments among original CPN, oscillation alleviation CPN and oscillation alleviation CPN with specific goal functions are examined.

\subsection{The Cognitive Packet Network}

The Cognitive Packet Network (CPN) is originally used for the large scale and fast-changing networks. As the conditions of networks typically change frequently, the protocols which periodically update the network status as well as the routing schemes that require the convergence of an entire network can always result in a constant lag and hinder the performance. Therefore, the concept of CPN is proposed to solve this problem by sending Cognitive Packets to monitor the network conditions and discover new routes sequentially. In the CPN, ``Cognitive packets'' play a dominating role in routing and flow control other than routers and protocols. Cognitive packets can adapt to pursue their predefined goals and learn from their own investigations and experience from other packets.
Each node of the CPN hosts a recurrent random neural network with a constant number of neurons (each neuron is associated with a neighbour node) and maintains a routing table which reserves a fixed number of QoS-oriented routes in a descending order. CPN carries three types of packets: smart packets (SPs), acknowledgements (ACKs) and dumb packets (DPs). SPs are used for information collection and routing discovery with respect to a specific QoS goal. They can either choose the most excited neuron as next hop or drift randomly to explore and measure undiscovered or not recently measured paths. When a SP reaches the destination, an ACK which stores all the gathered information will be generated and sent back to the source node through the reverse path. Because loops may be produced during route exploration, a loop remove algorithm is acted on the reverse path to take out any sequences of nodes with the same start and end node. When an ACK arrives at a node, it will update the routing table and train the random neural network by performing Reinforcement Learning (RL) \cite{gelenbe2001simulation}. DPs are the packets which actually carry the payloads. It always selects the best path at the top of the routing table as the next hop. When an ACK reach a node, it will trigger the learning process introduced in \cite{gelenbe1999cognitive,gelenbe2001measurement,gelenbe2004self,gelenbe2009steps} to calculate the excitement of each neuron.

According to Ref. \cite{gelenbe2004self}, the probability $q_i$ that a neuron is excited satisfies:

\begin{equation}
q_i = \left. \lambda^{+}(i) \middle/ [r(i) + \lambda^{-}(i)] \right.
\end{equation}
where
\begin{equation}
\lambda^{+}(i) = \sum_{j} q_jw^{+}_{ji} + \Lambda_{i} \quad \lambda^{-}(i) = \sum_{j} q_jw^{-}_{ji} + \lambda_{i}
\end{equation}

Here $w_{ji}^{+}$ is the probability that neuron $j$ emits positive impulse to neuron $i$ when $j$ is excited. $w_{ji}^{-}$ is the probability that neuron $j$ emits negative impulse to neuron $i$ when $j$ is excited. $r(i)$ is the total fire rate from neuron $i$. $\Lambda_{i}$ is the external arrival rate of positive signal to neuron $i$. $\lambda_{i}$ is the external arrival rate of negative signal to neuron $i$.

A goal function $G$ is used to compute a specific cost such as delay, jitter, energy consumption or a weighted combination from the source node to the destination node based on the information brought back by ACKs. A reward $R$ is then defined as $R = \frac{1}{G}$. The reward $R$ will be compared with a decision threshold $T_l$.

\begin{CJK*}{GBK}{song}
\begin{equation}
T_{l} = \begin{cases}0& \text{if it is the first time $R_l$ is obtained}\\aT_{l - 1} + (1 - a)R_{l}& \text{otherwise}\end{cases}
\end{equation}
\end{CJK*}
where $T_{l-1}$ is the previous value of the decision threshold $T_l$. Term $a$ is a constant $0 < a < 1$. $R_l$ is the most recently measured reward.

If $R_l$ is greater than or equals to $T_{l-1}$, it indicates the route from this neuron (node) satisfies the QoS required. Therefore the excitatory weight of this neuron should be strengthened significantly and the inhibitory weights of the other neurons should be increased slightly. Otherwise, we will punish the previous winner by increasing its inhibitory weight notably and raising the excitatory weights of the other neurons moderately.

 \begin{equation}
        \begin{aligned}
        \begin{split}
        & IF \quad T_{l-1}\leqslant R_l\\
        & w^{+}(i,j) \leftarrow w^{+}(i,j) + |R_l - T_{l-1}| \\
        & w^{-}(i,j) \leftarrow w^{-}(i,j) + \frac{|R_l - T_{l-1}|}{n-2}\quad \quad(i \not= j) \\
        & ELSE \\
        & w^{-}(i,j) \leftarrow w^{-}(i,j) + |R_l - T_{l-1}| \\
        & w^{+}(i,j) \leftarrow w^{+}(i,j) + \frac{|R_l - T_{l-1}|}{n-2}\quad \quad(i \not= j) \\
        \end{split}
        \end{aligned}
 \end{equation}
To ensure the summation of the probabilities that a neuron fires positive or negative impulses to all the linked neurons is 1, the total fire rate $r(i)$ from neuron $i$ is recorded before and after updating the fire rates.

\begin{equation}
r(i) = \sum_{m=1}^n[w^{+}(i,m) + w^{-}(i,m)]
\end{equation}

\begin{equation}
r(i)^{*} = \sum_{m=1}^n[w^{+}(i,m) + w^{-}(i,m)]
\end{equation}
where $n$ is the number of neighbour nodes.
Then the fire rates are re-normalized as follows:

\begin{equation}
w^{+}(i,j) \leftarrow w^{+}(i,j)*\frac{r(i)}{r(i)^*}
\end{equation}

\begin{equation}
w^{-}(i,j) \leftarrow w^{-}(i,j)*\frac{r(i)}{r(i)^*}
\end{equation}
Finally, the potential of a neuron can be calculated from equation (1) and (2).

In the context of emergency evacuation, because the environment is highly dynamic, SPs are used to search available routes and collect information such as the hazard intensity of a node. evacuees are considered as DPs and always follow the best path in the routing table. ACKs backtrack with collected information and update the excitement of traversed neurons (nodes). To apply the concept of CPN to emergency evacuation, the following requirements must be satisfied: a predefined graph-based layout of a confined place with information such as length of edges and node capacity is available. The edges of the graph depict the paths and the vertices represent the physical areas where a sensor is installed. The sensors could communicate with its neighbour nodes and detect the typical hazards such as fire, smoke, water, etc.

\subsection{Simulation model}

The Distributed Building Evacuation Simulator (DBES) is used to evaluate the performance of CPN as a decision support algorithm for emergency evacuation. DBES is a discrete event based simulation tool where each entity is modelled as a process. The simulator has a graphical user interface and includes different kinds of actors such as civilians, rescuers, firemen and evacuation wardens that can participate in the simulation as agents based on JADE \cite{jade}. The physical world in the simulator is represented by a set of ``Points of Interest'' (PoI) and the links between them.

\begin{figure}[!t]
\centering
\includegraphics[trim=30 30 30 30, clip, width=0.95\textwidth]{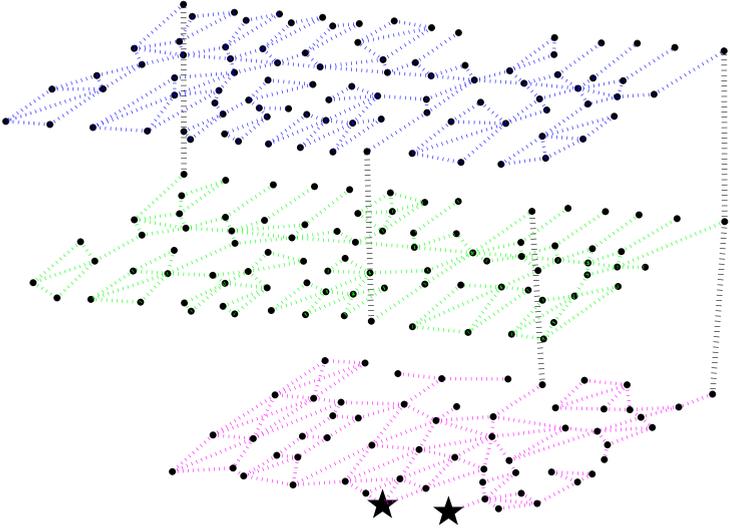}
\caption{The graph based layout of the building. The black stars represent the exits on the first floor.}
\label{fig: graph3D}
\end{figure}

The environment to egress is the lower floors of Imperial College's EEE building as shown in Fig. \ref{fig: graph3D}. SNs and DNs are deployed on the vertices of the environment. SNs collect hazard information and DNs execute the decision support algorithms. The goal function used is the effective path length which is proposed in \cite{filippoupolitis2009distributed}. Effective length combines the physical length of a path with the hazard intensity as shown in equation (9):

\begin{equation}
G_{d} = \sum_{i=1}^{n-1} E(\pi(i),\pi(i+1)) \cdot f(E(\pi(i),\pi(i+1)),t)
\end{equation}
where $\pi$ represents a particular path. Term $\pi(i)$ depicts the i-th node on the path $\pi$. Term $n$ is the number of nodes on path $\pi$. $E(\pi(i),\pi(i+1))$ is the physical length of the edge between node $\pi(i)$ and node $\pi(i+1)$ and $f(E(\pi(i),\pi(i+1)),t)$ is the fire intensity on the edge $E(\pi(i),\pi(i+1))$ at time $t$, so that:

\begin{equation}
f(E(\pi(i),\pi(i+1)),t) = \left\{
  \begin{array}{l l}
    >>\text{average edge length} & \quad \text{if the edge is affected by fire,}\\
    1 & \quad \text{otherwise.}
  \end{array} \right.
\end{equation}
This ensures the evacuees preferentially select paths without fire unless all paths have been exposed to hazard.

Three routing algorithms are employed in our experiment: ``autonomous'', Dijkstra's shortest path algorithm and CPN based algorithm. We assume that all the evacuees take a portable device with a predefined graph of the area in it and they can hear the fire alarm when the fire happens. ``Autonomous'' emulates the situation that evacuees do not receive any external assistance. Evacuees will simply follow the shortest path determined by their portable devices using Dijkstra's algorithm. When they encounter the fire, they will select an alternate route -- until they evacuate or perish. The second scenario simulates the cases where the evacuees receive assistance from decision nodes. When the fire spreads, the global hazard information will be synchronized in the network and the Dijkstra's shortest path algorithm will be executed at each decision nodes to obtain the up-to-date shortest path. The decision will be transformed to evacuees through visual indicators or portable devices. The third scenario is similar to the second one but using CPN based algorithm to make decisions.

\subsection{Results}

As there are several parameters in the CPN based algorithm that may affect the performance, we first test the CPN parameters before running full-scale simulations.

At the beginning of each simulation, before the fire happens, each node will have to send some SPs for routes discovery and also train the RNN. The detected paths and the knowledge of the RNN will be stored in the storage of nodes until a fire happens. We found that only 5 SPs can ensure at least one egress path is found. Meanwhile, 50\% of the nodes, especially the nodes on the main channels and the nodes near exits, can find the shortest path within 10 SPs. This is because a large amount of SPs sent from the whole network have transmitted through them. On the other hand, though the ``leaf'' nodes and the nodes far away from the exits may not find the shortest path within 10 SPs, the length of the shortest paths they found are below 120\% of the corresponding shortest path.

The results also indicates that the drift parameter significantly affects the paths found by CPN. The drift parameter is the probability that a smart packet follows RNN's advises (the most excited neuron) other than randomly chooses the next hop. If the drift parameter is too small, subsequent SPs will tend to follow the initial path returned by the first ACK. Because few new paths can be explored, the initial most excited neurons will be strengthened again and again. As the initial path is detected by SPs in a random manner, this route can be far below optimal. On the other hand, if the drift parameter is too large, the SPs tend to explore randomly in the network and the advantage of using CPN is weakened. Moreover, because there is a time-out threshold for paths stored in the routing table, the optimal path may be deleted due to it has not been visited for a long interval. However, a high drift parameter can discover much more paths than a low drift parameter. In our simulation, the initial drift parameter is set to 0.8 to guarantee the shortest paths can be found more quickly. When the fire starts, the drift parameter is set to 0.55.

To avoid overburdening the network, each SP has a limitation on the maximum hops that can be visited. As there are three floors in the building, we set a diverse maximum hops for each floor: the life-time constraint for the first floor is 60, for the second floor is 100 and for the third floor is 120. This setting ensures a detour path the can be found when the fire blocks a main channel.

\begin{figure}[!t]
\centering
\includegraphics[width=0.9\textwidth]{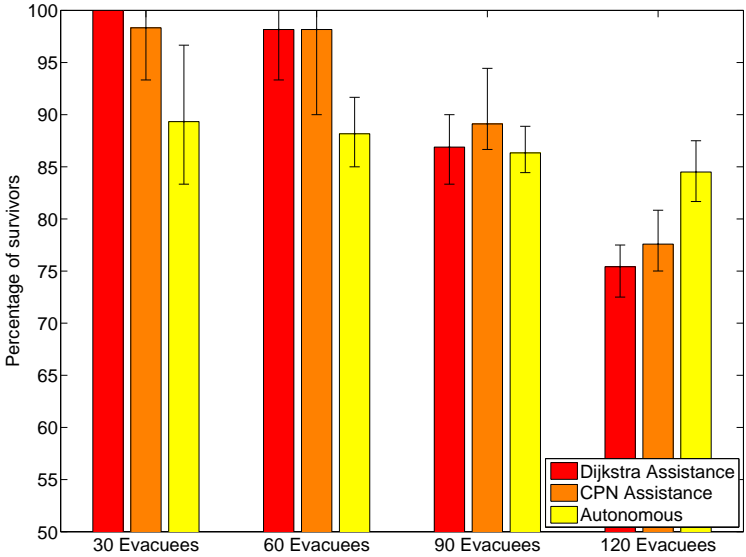}
\caption{Percentage of survivors on average for each scenario.}
\label{fig: simResults}
\end{figure}

Ten experiments where evacuees have randomized initial locations are conducted and results are shown in Fig. \ref{fig: simResults}. The histogram shows the percentage of survivors and the error bars presents single maximum and minimum value in the ten randomized experiments. It clearly shows that when the density of evacuees is relatively low, the CPN based algorithm reaches the performance of Dijkstra's shortest path algorithm. Meanwhile, roughly 10\% of ``autonomous'' evacuees die. This is because they follow the shortest path to the exits and only re-route when they encounter the fire. Because they make decision on the local hazard information and the fire spreads very quickly, ``autonomous'' evacuees may re-route several times and cause a long time delay. While they backtrack, the fire may have already blocked all the main routes and trapped them in the building. Furthermore, CPN based algorithm and Dijkstra's shortest path algorithm have a vision of the whole area or a set of paths respectively and can provide early warning and re-route evacuees before they reach the hazard. As the intensity of evacuees increases, the performance of the assistance system progressively reduces. When there are 90 evacuees in the building, CPN based algorithm performs the best and Dijkstra's shortest path algorithm and autonomous algorithm have approximately the same performance. When there are 120 evacuees in the building, the autonomous algorithm performs the best and the assistance systems become detrimental to the evacuees. This is largely due to congestion as shown in Table 1. Dijkstra's shortest path algorithm tends to direct all the evacuees to a single optimal route. Hence, numerous congestion and stampedes will be created due to the insufficiency of the path capacity. CPN based algorithm also tends to guide evacuees to the shortest path. However, as a decentralized algorithm, the nodes do not update the paths and the hazard information at the same time. In addition, as the hazard spread continuously, some nodes may not always resolve the shortest path but an approximate shortest path. The above natures of CPN tend to distribute evacuees and in turn ease congestion to some extent.

\begin{table}
    \begin{center}
        \begin{tabular}{| r | c | c | c |}
            \hline
            Number of Evacuees & Autonomous Paths & CPN Based Paths & Dijkstra Based Paths \\ \hline
            30 ~~~~~~~~ & 89.6 & 104.8 & 105.2 \\ \hline
            60 ~~~~~~~~ & 308.2 & 373.4 & 389.2 \\ \hline
            90 ~~~~~~~~ & 557.2 & 632.6 & 671 \\ \hline
            120 ~~~~~~~ & 833.6 & 904.8 & 951.8 \\ \hline
            \hline
        \end{tabular}
        \caption{The amount of congestion happens on each navigation algorithms on average.}
        \label{table: nonEmptyNodes}
    \end{center}
\end{table}

Different from Dijkstria's shortest path algorithm which only re-routes when the shortest path is affected by fire, the CPN based algorithm can switch before the hazard reaches. This is because the SPs do not only transmit through the shortest path, but also drift alongside it. Once a drifting SP is exposed to the fire, it will return a much larger effective length and the related neurons will be punished. After a while, the most excited neuron of the corresponding node will change and new paths away from the hazard will be constructed subsequently.

\subsection{Routing diverse categories of evacuees}

Research on emergency evacuation has previously focused on ``normal'' individuals with full mobility as well as first responders and security personnel. However, there are commonly diverse categories of evacuees such as adults, children, people in wheelchairs, ill people, etc. Therefore, it is more optimal to satisfy their specific QoS criteria, such as time consumption, power utilization, security, etc. One advantage of the CPN based algorithm is that it can fulfil diverse criteria because each SP can explore the environment based on its specific QoS need. In this section, we present three different goal functions to fulfil the specific needs of diverse evacuees.

\subsubsection{Congestion management}

Most of the current evacuation routing algorithms do not take congestion into account. They either direct the evacuees to the safest paths or to the shortest paths. Ref. \cite{chen2008load,chen2011distributed} propose a distributed navigation protocol to guide evacuees to exits with minimal congestion. Each sensor is assigned a potential which is determined by the number of evacuees in its vicinity and paths are constructed by the potential field. However, the spread of hazard is not considered and therefore evacuees can be directed to a hazardous area. Moveover, due to the movement of evacuees are highly dynamic, this protocol may induce ``oscillation'' \cite{gelenbe2009steps}. EVACNET+ \cite{chalmet1982network,francis1984negative,kisko1985evacnet+} converts the evacuation network to a time-expended network to reserve edge and node capacity to evacuee flows with regard to time series. However, the computing load is extremely high and can not handle dynamic hazard. Moreover, the requirement of precise expected arrival and departure instants for an evacuee to each edge in a complex environment is unrealistic. M/M/1 queueing model \cite{gelenbe1987introduction} is employed in Ref. \cite{desmet2013graph} to predict the location of congestion. However, the current congestion located on a path with long duration may influence evacuees move towards it. Therefore, we propose a improved algorithm based on M/M/1 queueing model to predict the congestion that happens on each node with regard to current congestion.

As a vital component of evacuation models, analytical model has been long applied in this area. Ref. \cite{macgregor1991state} applies state-dependent queueing models to analyse the emergency evacuation planning (EEP) problem. Ref. \cite{bakuli1996resource} uses M/G/C/C model along with the Mean Value Analysis (MVA) algorithm to resize the width of the passageways. An undirected graph model is used in \cite{fiedrich2000optimized} to represent the disaster area after earthquake and heuristic procedures are used to optimise the allocation of resources. Ref. \cite{hasofer2001stochastic} uses two networks to model the spread of a fire and the egress process of civilians respectively. A discrete hazard function is employed to simulate the fire spread stochastically. A probability based approach is proposed in Ref. \cite{elms1984modeling} to evaluate the fire-spreading in a building and the effect of door fire-resistance ratings along with the presence of sprinklers are investigated. Ref. \cite{desmet2013graph} presents the formula of average traversal time of an edge based on the state-steady of a Jackson network model. The authors assume that the traversal time of a single edge is much shorter than the overall evacuation time, the arrival of evacuees is subject to Poisson distribution and the service time yields exponential distribution.

The evacuation process can be considered as a steady-state because the total evacuation time is much longer than the time to traverse any single edge \cite{desmet2013graph}. Therefore, according to \cite{gelenbe1987introduction}, the average queue length $L_{queue}$ of a node in a steady-state is:
\begin{equation}
L_{queue} = \frac{\rho}{1-\rho}
\end{equation}
where $\rho = \frac{\lambda}{\mu}$ is the probability that a node has at least one evacuee, $\lambda$ presents the Poisson arrival rate, $\mu$ depicts the exponential service rate. If $\rho >= 1$, then we suppose congestion will happen when a new evacuee reaches.

The arrival rate $\lambda$ is recorded by each node in a real time manner. To reflect the history situation and a possible queue of a node, we make use of a rolling average mechanism other than use the latest obtained arrival rate directly.
\begin{equation}
R_c = aR_h + (1-a)(\frac{1}{T_c - T_h})
\end{equation}
where $a$ is a constant $0 < a < 1$. $R_c$ is the current arrival rate. $R_h$ is the previous arrival rate. $T_c$ is the current instant when an evacuee arrives. $T_h$ is the previous instant when an evacuee arrived.

The service rate $\mu$ is inversely proportional to the service time of a node, which is the interval for an evacuee to traverse the node. In the current DBES, the service rate is 1.

Long duration congestion that happens on a certain node may also affect an evacuee because it may still remain there when the evacuee reaches that node. Hence the current congestion on a path should also be taken into account. This algorithm evaluates if congestion located on a certain node will affect an evacuee by comparing congestion lasting interval and the arrival time of the evacuee. The details of the approach to predict the number of congestion that will be encountered by an evacuee on a path is shown in \textbf{Algorithm 1}.
\begin{algorithm}[!htb]
\caption{Predict the potential amount of congestion encountered by an evacuee on a path}
\textbf{Data:}\ A path explored by SPs from a node\\
\textbf{Result:}\ The potential number of congestion encountered by an evacuee traversing across the path
\begin{algorithmic}[1]
\STATE Set the total congestion number $C_{total}$ to 0
\STATE Set the total travel time $t_{total}$ to 0
\FORALL {the edges $e$ on the path}
\STATE Set the time cost on the edge $t_{edge}$ to 0
\STATE Calculate $t_{edge}$, $t_{edge} = \frac{E_{edgelength}}{V_{speed}}$, where $E_{edgelength}$ is the length of $e$ and $V_{speed}$ is the speed of the civilian.
\STATE /*Add the predicted number of congestion on the source node of the edge $e$*/
\STATE $C_{node} = C_{total} + \rho$
\STATE /*Add the time cost on the source node of the edge $e$*/
\STATE $t_{node} = a \cdot \frac{\rho}{1-\rho}$, where $a$ is the coefficient between potential queue length and the waiting time
\STATE $t_{edge} = t_{edge} + t_{node}$
\STATE /*Determine if the current congestion located on this edge will affect the evacuee*/
\IF {the current queue length $n$ on the source node of the edge $e$ is not zero}
\STATE Calculate the queueing time $t_{queue} = b \cdot n$, where $b$ is the coefficient between current queue length and the waiting time
\IF {$t_{queue} > t_{total}$}
\STATE /*This means the congestion will affect the evacuee*/
\STATE $C_{total} = C_{total} + 1$
\STATE $t_{edge} = t_{edge} + t_{queue} - t_{total}$
\ENDIF
\ENDIF
\STATE $t_{total} = t_{total} + t_{edge}$
\ENDFOR
\STATE \textbf{Return} $C_{total}$
\end{algorithmic}
\end{algorithm}
\subsubsection{Oscillation Alleviation}

Oscillation will occur when a sensible metric \cite{gelenbe2003sensible} is used or the process of updating sensor readings and arrival of ACKs are asynchronous when the hazard spreading quickly. A detailed literature review is presented in \cite{gelenbe2009steps} with several resolutions for oscillation in packet networks: set a path switch probability; arrange a QoS gain threshold to switch routes until the improvement has reached a certain value; ensure the usage of a path exceed a minimum number of packets.

In this section we use a method similar to the last approach above to mitigate oscillation. ``Movement depth'' is introduced to ensure the evacuees could only accept new suggestion after traversing a certain number of nodes. ``Movement depth'' is the number of hops an evacuee should traverse before switch.

\begin{figure}[htb]
\centering
\begin{tabular}{ccc}
\bmvaHangBox{\fbox{\includegraphics[width=3.7cm]{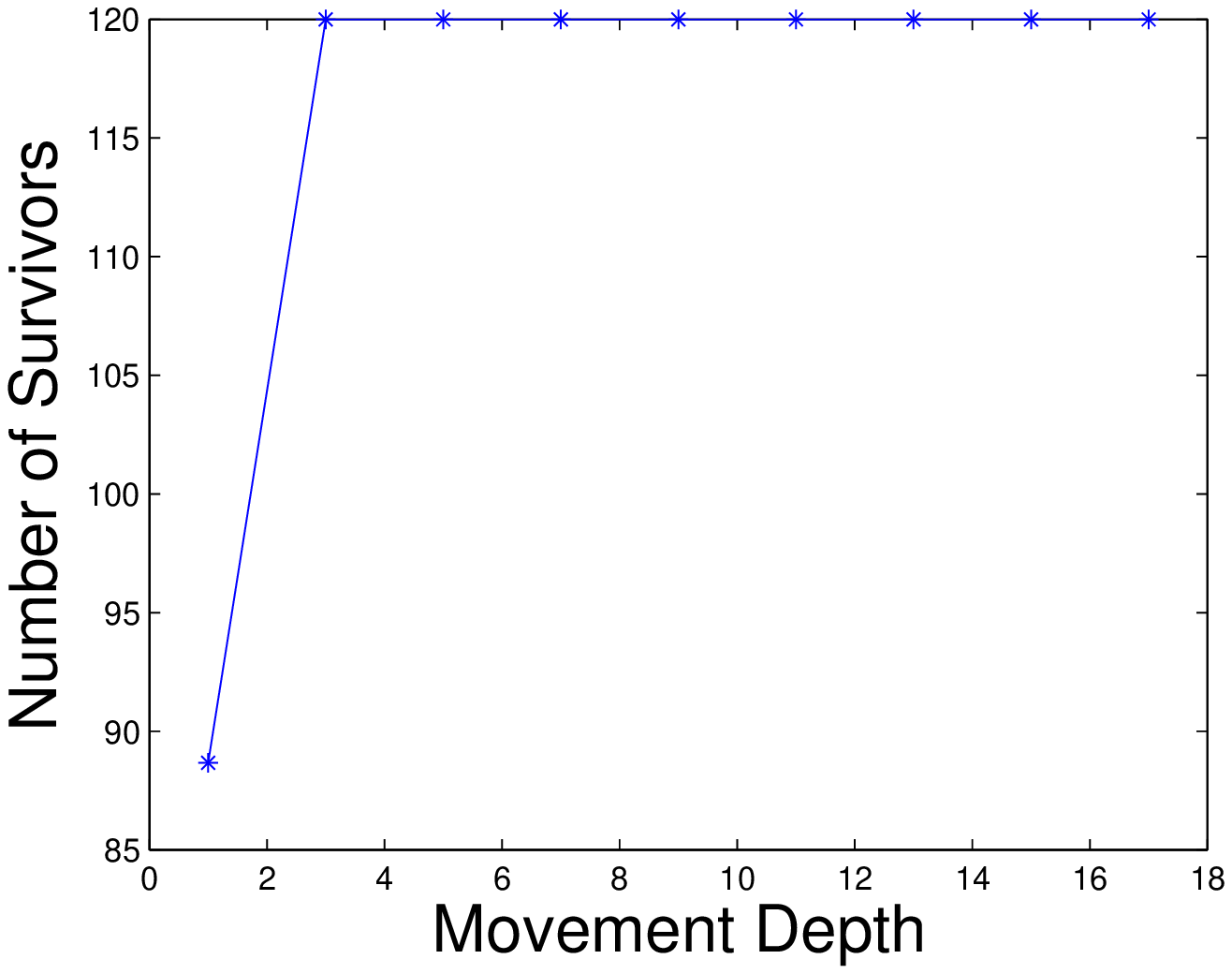}}}&
\bmvaHangBox{\fbox{\includegraphics[width=3.7cm]{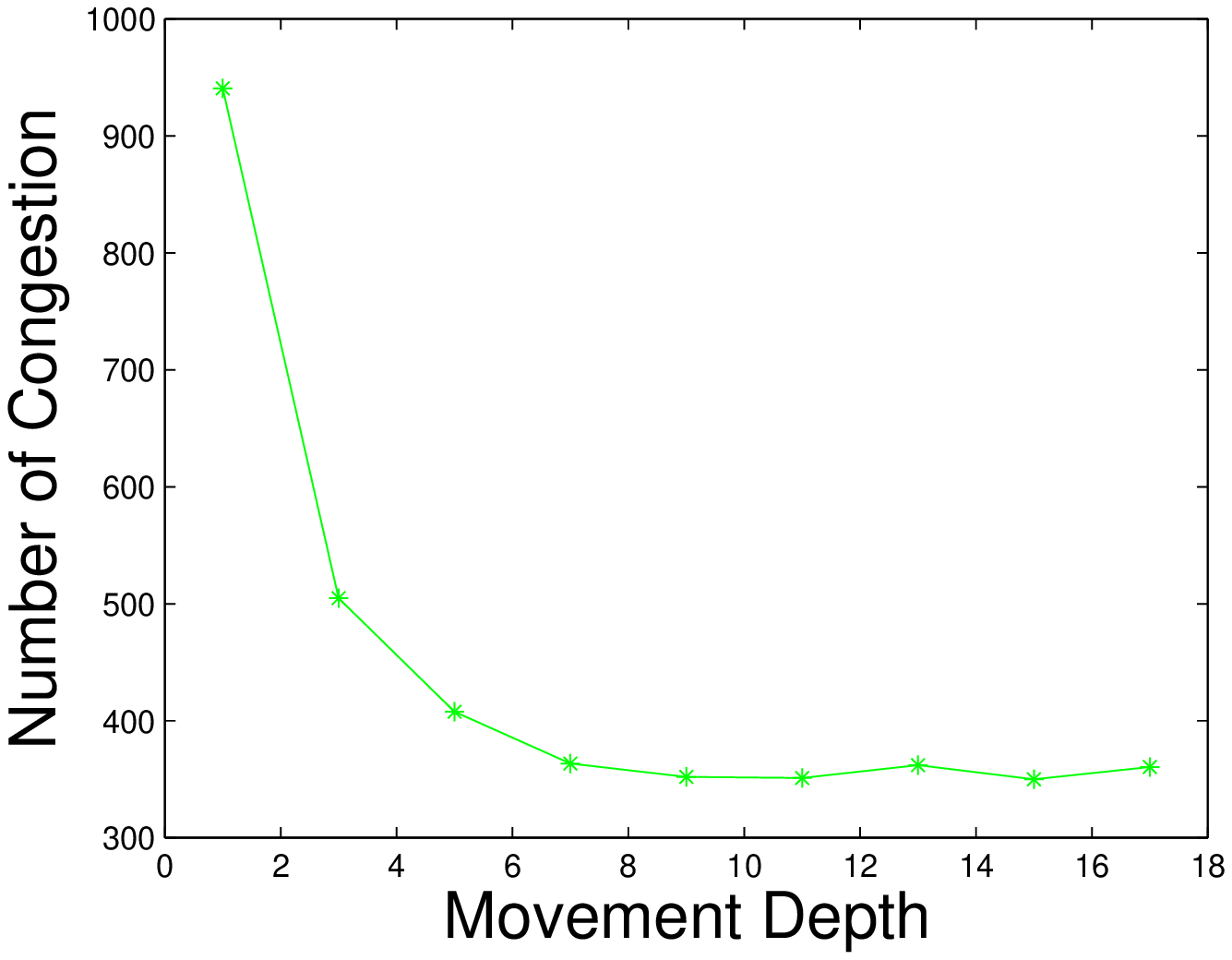}}}&
\bmvaHangBox{\fbox{\includegraphics[width=3.7cm]{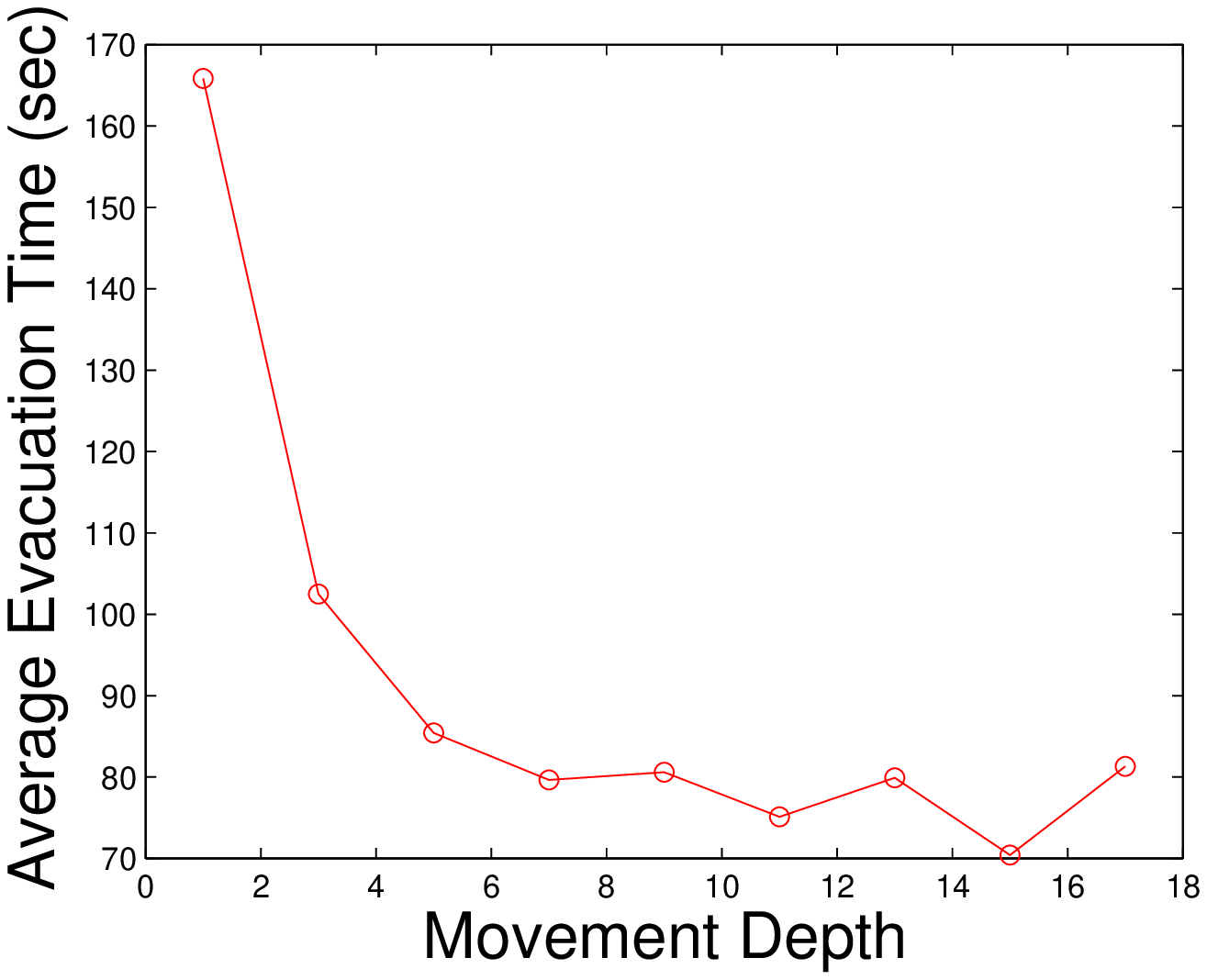}}}\\
(a)&(b)&(c)
\end{tabular}
\caption{Figure (a) is the number of survivors for different values of movement depths. Figure (b) is the amount of congestion for different values of movement depths. Figure (c) is the average evacuation time for different values of movement depths.}
\label{fig:teaser}
\end{figure}

Three randomized iterations have been conducted for movement depth from 1 to 17 in a scenario with 120 evacuees. We use this scenario because the effect of oscillation is more obvious in high density environments. Fig. \ref{fig:teaser} shows that there is a notable rise on the algorithm's performance when the movement depth increases from 1 to 3. However, the performance becomes stable when the movement depth reaches 5. Fig. \ref{fig:teaser} (a) indicates that when the variable is 1, approximately 27\% of the evacuees will die while no evacuee will perish if the movement depth is larger than or equals to 3. This is because CPN based algorithm evaluates the QoS metric of a whole path other than a single edge. In other words, the evacuees should finish the whole path to satisfy their QoS needs if the condition does not change. If they continuously change paths, they will not follow the optimal path but a collection of the first edge of optimal paths. In the context of emergency evacuation, oscillations are mainly caused by the spreading of the hazard. Fig. \ref{fig:teaser} (b) and (c) show that mitigating oscillation will significantly reduce congestion and therefore decrease average evacuation time markedly.

\subsubsection{Specific Goal Functions}

The typical metric used in CPN based algorithm is the distance-oriented goal function, which pursues the shortest path from a source to a destination. To fulfil the specific QoS needs of diverse categories of evacuees, three kinds of novel metrics, namely time-oriented goal function, energy efficiency oriented goal function and safety-oriented goal function are presented.

The time-oriented goal function pursues the shortest time cost to reach an exit. It suffices the QoS need of normal evacuees.
\begin{equation}
G_{t} = \sum_{i=1}^{n-1}\left\{\frac{E^{e}(\pi(i),\pi(i+1))}{V_{speed}}+a\cdot\frac{\rho_{\pi(i)}}{1-\rho_{\pi(i)}}+ K[b\cdot C_{\pi(i)}-t_{total}^{\pi(i+1)}]\right\}
\end{equation}
where $\pi$ represents a particular path. Term $n$ is the number of nodes on path $\pi$. Term $\pi(i)$ depicts the i-th node on the path $\pi$. $E^{e}(\pi(i),\pi(i+1))$ is the effective length of the edge between node $\pi(i)$ and node $\pi(i+1)$. $V_{speed}$ is the average speed of this category of civilians. Term $a$ is the coefficient between waiting time and the potential queue length. $\frac{\rho_{\pi(i)}}{1-\rho_{\pi(i)}}$ is the predicted queue length of node $\pi(i)$. $K[X]$ is a function that takes the value X or 0 if X is larger than 0 or 0 if X is smaller than 0 respectively. $C_{\pi(i)}$ is the current queue length of node $\pi(i)$. Term $b$ is the coefficient between waiting time and the current queue length. In this experiment, $a = b = 1$. $t_{total}^{\pi(i+1)}$ is the overall time for the evacuee to reach node $\pi(i+1)$, it can be calculated by \textbf{Algorithm 1}.

The energy efficiency oriented goal function search the path with minimum energy consumption. This function can be used for wheelchairs or robots due to the battery power limitations.
\begin{equation}
G_{e}=c_1\cdot C_{total}+\sum_{i=1}^{n-1}c_2\cdot E(\pi(i),\pi(i+1))+\sum_{i=2}^{n-1}c_3\cdot\theta (\pi(i-1),\pi(i),\pi(i+1))
\end{equation}
where $\pi$ represents a particular path. Term $n$ is the number of nodes on path $\pi$. Term $\pi(i)$ depicts the i-th node on the path $\pi$. $C_{total}$ is the total number of congestion calculated by \textbf{Algorithm 1}. Term $c_1$ is the energy consumption of a wheelchair when a breaking event happens, $c_2$ is the energy consumption per centimetre when moving straight on an edge. $E(\pi(i),\pi(i+1))$ is the physical length of the edge between node $\pi(i)$ and node $\pi(i+1)$. Term $c_3$ is the energy consumption per degree when a turning event happens, $\theta (\pi(i-1),\pi(i),\pi(i+1))$ is the angle between edge $E(\pi(i-1),\pi(i))$ and $E(\pi(i),\pi(i+1))$.

The safety-oriented goal function discovers the safest path. It can be used for sick people or children as they are more likely to faint due to the impact of the poisonous smoke in a fire. To estimate the hazard intensity of a node, we assume the hazard spread rate is $a$ (cm/s) and the hazard growth rate at a node is $b$.

\begin{equation}
G_{s} = \sum_{i=1}^{n-1}1[t_{evacuee}^{\pi(i+1)}+t_{current}<t_{hr}^{\pi(i+1)}]\cdot b \cdot (t_{evacuee}^{\pi(i+1)}+t_{current}-t_{hr}^{\pi(i+1)})+E^{e}(\pi(i),\pi(i+1))
\end{equation}
where $\pi$ represents a particular path, $n$ is the number of nodes on path $\pi$, $\pi(i)$ depicts the i-th node on the path $\pi$. $1[X]$ is a function that takes the value 1 or 0 if X is false or true respectively. The term $t_{evacuee}^{\pi(i+1)}$ is the time cost for an evacuee to reach node $\pi(i+1)$ with regard to congestion. The term $t_{current}$ is the current simulation time. The term $t_{hr}^{\pi(i+1)}$ is the time cost for the hazard to reach node $\pi(i+1)$. $E^{e}(\pi(i),\pi(i+1))$ is the effective length between node $\pi(i)$ and node $\pi(i+1)$. We employ $E^{e}(\pi(i),\pi(i+1))$ to ensure the value of $G_{s}$ is not nil if the hazard does not reach the path during a civilian's evacuation process.

The details of this algorithm is shown in \textbf{Algorithm 2}.
\begin{algorithm}
\caption{Predict the potential hazard of a path}
\textbf{Data:} a Path explored by SPs from a node\\
\textbf{Result:} the potential hazard of the path
\begin{algorithmic}[1]
\FORALL {the nodes $n$ in the environment}
\STATE Find the shortest path $D$ from $n$ to the fire node by using the built-in map
\STATE Calculate the time $t_{hr}$ cost for the hazard to reach $n$
\STATE $t_{hr} = D/a$
\ENDFOR
\STATE \textbf{Towards} each path in the routing table at a certain node $n_{source}$
\STATE Set the total potential hazard $H_{total}$ of a path to 0
\FORALL {the nodes $n_p$ $\in$ Path $\pi$}
\STATE /*Calculate the time cost $t_{evacuee}$ for an evacuee to reach $n_p$ from $n_{source}$*/
\STATE $t_{evacuee}$ can be calculated by \textbf{Algorithm 1}
\STATE /*Calculate the potential hazard $h_{node}$ when the evacuee arrives*/
\STATE $h_{node} = b\cdot (t_{evacuee}+t_{current}-t_{hr}) + E(\pi(i),\pi(i+1))$
\STATE $H_{total} = H_{total} + h_{node}$
\ENDFOR
\STATE \textbf{Return} $H_{total}$
\end{algorithmic}
\end{algorithm}
\subsubsection{Experimental Results}

The experiments that feature the three goal functions above are conducted on scenarios with 30, 60, 90, 120 evacuees respectively. The original distance-oriented goal function based experiments are also run for comparison purpose.

Three scenarios are employed in this experiment. The first scenario is the original CPN based algorithm with the distance-oriented metric. It has no specific measures to ease oscillation since the movement depth is set to 1. The second scenario is similar to the first one but the movement depth is 3. The aim is to maintain the sensitivity and adaptivity of the CPN based algorithm. The reason is that if the movement depth is too large, the evacuees cannot react fast enough to an ongoing hazard. The movement depth of the third scenario is also 3 but with different kinds of goal functions.

\begin{figure}[htb]
\centering
\includegraphics[trim=30 30 30 30, clip, width=0.95\textwidth]{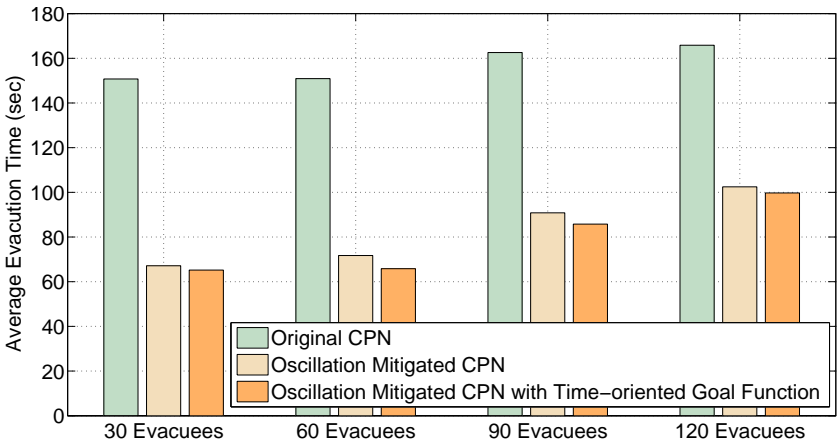}
\caption{Average evacuation times for three scenarios.}
\label{fig: timeoriented}
\end{figure}

\begin{figure}[htb]
\centering
\includegraphics[trim=30 30 30 30, clip, width=0.95\textwidth]{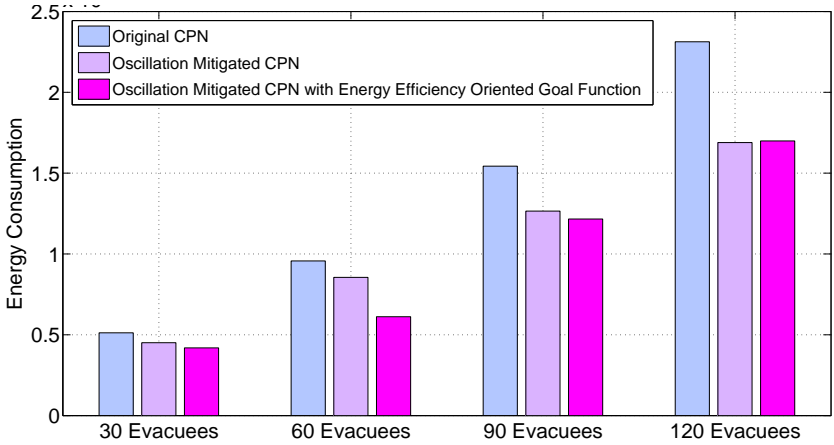}
\caption{Average energy consumption for three scenarios.}
\label{fig: energyoriented}
\end{figure}

\begin{figure}[htb]
\centering
\includegraphics[trim=30 30 30 30, clip, width=0.95\textwidth]{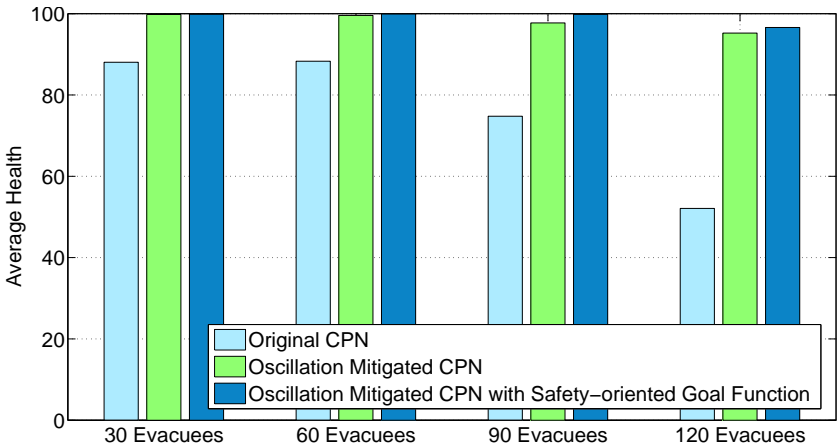}
\caption{Average health for three scenarios.}
\label{fig: safetyoriented}
\end{figure}

The results in Fig. \ref{fig: timeoriented} shows that scenario 2 and 3 could both reduce the average evacuation time significantly by easing oscillation. Scenario 3 performs better than scenario 2 because the time-oriented goal function is used. The results also prove that the congestion management algorithm in Section 3.5.1 is effective. The power utilization shown in Fig. \ref{fig: energyoriented} are calculated by considering breaking events (cost 50 units of energy), turning (cost 2 units of energy per degree), and moving forward (cost 1 unit of energy per centimeter). It reveals interesting results. Scenario 2 and 3 always perform better than scenario 1 as oscillation alleviation can reduce congestion. Hence, the breaking events are decreased and less energy is cost. However, comparing with scenario 2 and 3, the current energy efficiency goal function can only reduce energy consumption in environment with relatively low population density. The saved energy is 31796.3709, 243531.1615 and 49019.346 units for 30, 60, 90 evacuees respectively. The distance-oriented goal function performs slightly better than energy efficiency-oriented goal function when there are 120 evacuees. This is because the energy efficiency goal function is a more sensitive metric (each congestion consume 50 units of energy) than distance-oriented goal function and it will be affected by oscillation more seriously. Fig. \ref{fig: safetyoriented} shows the average health of evacuees. We set a high initial fire intensity to make the results more obvious. The results indicate that the safety-oriented algorithm always performs the best among the three algorithms. That is because this algorithm not only consider a node's current fire intensity but the situation when it reaches that node. Actually, the safety-oriented goal function is a tradeoff between the number of congestion and the frequency of switching directions. The safety-oriented goal function generates more congestion than the distance-oriented goal function because it selects the future safest path other than the current safest path. However, the safety-oriented goal function assisted evacuees tend to have lower probabilities to be blocked by the hazard and therefore choose an alternate path.

\section{Research Plan}

Table \ref{table: future} gives an overview of the key research milestones throughout my PhD. At present, I have completed the first two milestones. A brief overview of the other milestones is given below, highlighting the expected original contributions and the tasks involved.

\begin{table}
\begin{center}
\begin{tabular}{|p{8cm}|c|c|}
\hline
Research Milestone & Target Date & Status \\
\hline
Apply CPN to emergency evacuation & 10/04/2013 & Completed \\
\hline
Investigate appropriate metrics for diverse categories of evacuees & 10/07/2013 & Completed \\
\hline
Apply G-networks models to emergency evacuation & 10/10/2013 & In Progress \\
\hline
Investigate novel algorithms for route searching & 10/01/2014 & Not Started \\
\hline
Investigate animal foraging inspired optimization algorithms & 10/05/2014 & Not Started \\
\hline
Reduce computational complexity of decision-making in task assignment & 10/09/2014 & Not Started \\
\hline
Use virtual or augmented reality to facilitate decision making & 10/03/2015 & Not Started \\
\hline
Improve security of network based emergency evacuation systems & 10/07/2015 & Not Started \\
\hline
Exploit decision support algorithms in degraded conditions & 10/11/2015 & Not Started \\
\hline
Realise energy efficiency in power distribution in emergency & 10/03/2016 & Not Started \\
\hline
First draft of thesis & 10/10/2016 & Not Started \\
\hline
\end{tabular}
\label{table: future}
\end{center}
\caption{Summary of research milestones throughout the duration of the PhD.}
\end{table}

\vspace{0.5em}
\noindent\textbf{Reduce computational complexity of decision-making in task assignment}
\vspace{0.5em}

Emergency management can be modelled as nonlinear combinatorial optimisation problems to assign resources to tasks while minimizing the overall cost \cite{gelenbe2010randomemergency}. Due to the fact that classical optimization algorithms can be easily trapped in local minima \cite{samadzadeganbiologically} and exact solutions can be computationally expensive and time consuming for such NP-hard problems \cite{ahuja2007exact}, meta-heuristics algorithms inspired by natural and artificial intelligence have motivated considerable research.

A survey of the literature indicates rare research has discussed the resource allocation problem for emergency. Ref. \cite{fiedrich2000optimized} employs both simulated annealing algorithm and tabu search algorithm to allocate rescue machines and equipment to different operational areas namely disaster areas, stabilising areas and immediate rehabilitation areas. Ref. \cite{gelenbe2010randomemergency} uses RNN to assign ambulances (assets) to injured individuals (tasks). This is a semi-assignment problem because each task can be allocated to multiple assets while each asset can only be assigned to one task. Each task-asset pair is viewed as a neuron in the RNN model. The neurons are fully connected because each task can be executed by any one of the resources. A neuron can send positive or negative impulses to linked neurons and the unique solution can be obtained from neurons with highest excitement in real time manner.

Although heuristic algorithms can not guarantee to discover the optimal solution but they can find suboptimal solution in polynomial time. Ref. \cite{aguilar1997task} investigates the problem of assigning tasks to processors of a parallel program to minimise a compound goal function related to cost of communication, access files, interference and load balance. The random neural network model, simulated annealing, genetic algorithm and well-known Kernighan-Lin graph-partitioning greedy heuristic \cite{kernighan1970eflicient} are executed with the purpose of comparison. The results show that the former three algorithms perform better than Kernighan-Lin heuristic. Simulated annealing provides best results while the genetic algorithm performs the second to best. However, these two algorithms are very time-consuming compared with RNN and Kernighan-Lin heuristic. Kernighan-Lin heuristic runs very fast but with worst results. Compared with the other three algorithms, RNN gives an equilibrium result and it is easy to be deployed on a distributed system.

According to the above results, we will employ heuristic algorithms such as RNN or scheduling algorithms based on queueing models \cite{chabridon1997scheduling} to reduce the complexity of decision-making in emergency management. For instance, it is a dynamic task assignment problem for fireman to rescue injured civilians in a burning building. Another interesting problem worth to be studied is that we found decision nodes on main channels suffer from high computational load while nodes at less strategic positions (e.g. a room) are idle most of the time. This will drain the battery power of decision nodes at strategic location rapidly and shorten the system lifetime. Therefore, we will exploit load-balancing algorithms for task migration which is a mechanism to transfer a process from one node to another at the cost of additional communication overhead.

\vspace{0.5em}
\noindent\textbf{Investigate animal foraging inspired optimization algorithms}
\vspace{0.5em}

Millions of years' evolution has made the animal foraging behaviours become near-optimal solutions of autonomous search. These behaviours are commonly constructed on top of simple but reliable mechanisms and therefore suit the contexts with limited computational power. Ref. \cite{gelenbe1997autonomous} presents a comprehensive literature review of biomimetic models based on animal research. Inspired by this work, we will investigate efficient algorithms and policies for searching paths and task assignment.

American white pelicans increase degree of coordination when the prey capture rates decrease \cite{mcmahon1992foraging}. This foraging strategy that alerts with prey capture rates can benefit the exploration process of SPs. For instance, when ACKs tend to bring back new paths (high capture rates), the drift parameter will be set to a higher value adaptively to disseminate SPs more randomly. In contrary, when ACKs are prone to bring back paths that are already stored in the routing table, the drift parameter will be automatically assigned to a low value and SPs will concentrate on monitoring the current paths.

Ref. \cite{noda1994local} investigates the prey searching behaviours of stout-body chromis (Ss). They usually search for zooplankton in a tortuous pattern within the conventional foraging regions, regardless of presence or absence of prey. In contrast, Ss will move rapidly between the foraging regions without searching. Similarly, each node in the CPN should be able to adapt its drift parameter in accordance with its position other than set equivalent quantities. A node in a narrow corridor should have a smaller drift parameter for quick path-finding and rapid adjustment to a spreading hazard while a node at a hall should have a larger drift parameter to discover more paths.

According to their ages, wild female pipevine swallowtail butterflies select different shapes of leaf as hosts \cite{papaj1986shifts}. Similarly, by containing information such as private cognitive map and executable code \cite{gelenbe1999cognitive}, a portion of SPs can seek to avoid the existing paths in the routing table and explore new paths more efficiently. On the other hand, inspired by the foraging behaviours of ant or bee colony, when the QoS of a returned route is the current best in the routing table, some source routed SPs can be generated to follow this route and accelerate the self-adaptive process of CPN. This can be particularly effective to a fast-changing hazardous area where the best route can change suddenly. One advantage of this approach is that the learning rate $a$ in equation (3) is not changed because a large $a$ can slow down the self-adjustment process while a small $a$ will induce the network unstable.

The searching algorithms for the rescuers to find injured civilians can also be inspired by certain animal forging strategies. In attempting to find more food, the beetle Gyrinus picipes will significantly reduce the step length, linear displacement and path straightness while increase thoroughness and turning angle after a prey is captured \cite{winkelman1991gyrinid}. By examining a hypothetical forage searching for immobile, random distributed prey from a central position and homing straight back, an ``optimal sinuosity'' formula is proposed in \cite{bovet1991optimal} to determine the optimal sinuosity of a search path which minimises the total distance of exploring a given area.

\vspace{0.5em}
\noindent\textbf{Investigate novel algorithms for route searching}
\vspace{0.5em}

One of the disadvantage of the current RNN based CPN is that the probability for a neighbour node to be selected as the next hop is equivalent except the node with highest excitement. Due to the randomness, it may cost more time for SPs to find the best path as well as hinder the learning efficiency of RNN (Due to the frequent punishment on the most excited neuron for finding a bad path caused by random walk). Since CPN is a open network protocol that can combine with various decision making algorithms \cite{filippoupolitis2010emergency}, we will investigate other algorithms to discover best path more efficiently in the future research.

Ref. \cite{gelenbe1998autonomous} presents a simplified infinite horizon optimization (SIHO) algorithm to guide robots for demining. To make use of global information other than move based on local perception, SIHO algorithm use gradient descent to optimise the probabilities of moving to neighbouring points. The results show that SIHO can maximise the overall probabilities of finding mines at all positions. Similarly, a novel QoS-driven sensible routing policy is proposed in Ref. \cite{gelenbe2003sensible} to determine the probabilistic choices among all possible paths based on expected value of QoS metrics. The probability of selecting a path is inversely proportional to the QoS of the path. The work also proves that a m+1-sensible policy which uses the (m+1)-th power of the QoS for each alternate path performs better than m-sensible routing policy. Inspired by the above work, we will exploit algorithms to optimise the probabilistic choices of all neighbour nodes to increase the efficiency of finding the best path.

In contrary to the conventional discrete event simulations which enumerate all possible behaviours that may arise, Ref. \cite{gelenbe2001simulation} use learning agents to model the adaptive behaviour of human and pursue certain goals autonomously and adaptively. Each learning agent has a internal representation that is constructed by its own observation of the simulation environment and experience from other agents. Three learning paradigms are used for learning agents to make decisions based on the internal representation: adaptive finite-state machines, feedforward random neural network learning and reinforcement random neural network learning. Motivated by this paper, we will seek to use adaptive stochastic finite-state machines (ASFSMs) \cite{viswanathan1972comparison,narendra1983use} as well as feedforward random neural networks to emergency evacuation routing and examine their performance.

To improve the path searching efficiency of CPN, genetic algorithms \cite{gelenbe1996genetic} will also be exploited to generate new paths by splicing existing routes. As the QoS metrics for all categories of evacuees are additive, we can employ the genetic algorithm presented in \cite{gelenbe2006genetic} to create valid paths. The genetic algorithm will be executed on each decision node. The paths discovered by SPs are viewed as initial chromosomes which can evolve under the rules as follows (Each path presents a type of chromosome).

\begin{itemize}
\setlength{\itemsep}{1pt}
\setlength{\parskip}{0pt}
\setlength{\parsep}{0pt}
\item A chromosome of type $i$ can be automatically generated.
\item A chromosome of type $i$ can survive or perish in accordance with its ``fitness''. The ``fitness'' attribute can be evaluated by the QoS of this path.
\item A chromosome of type $i$ can mutate into a chromosome of type $j$.
\item A chromosome of type $i$ can combine with an identical or different type of chromosome to produce a new genotype when the parental chromosomes (paths) have the same intermediate node. In this case, the originating chromosomes will be replaced by the new genotype (the new genotype can be the same as the parental chromosomes).
\item A chromosome of type $i$ can combine with a chromosome of type $j$ and both perish.
\end{itemize}

One advantage is that long term behaviour of this model can be obtained by a formula of the joint steady-state probability of the number of chromosomes of each type other than long run simulation. The author has proved that: (1) the steady-state probability of the number of chromosomes of each type is the product of the marginal probabilities for each chromosome type and independent from the initial state configuration; (2) the marginal probability of each type of chromosome yields a geometric distribution. Due to the closed solution of the model, gradient based algorithms can be used to minimise diverse goal functions by optimising the parameters corresponding to fitness, mutation, crossover, etc.

\vspace{0.5em}
\noindent\textbf{Use virtual or augmented reality to facilitate decision making}
\vspace{0.5em}

Compared with conventional emergency response systems with limited computational power, recently developed systems based on grid computing and cloud computing can assimilate multi-domain sensor information, store a huge amount of data and provide powerful computing capacity. As a result, virtual reality or augmented reality can be adopted on smart handsets for emergency routing couched on the data from cloud severs \cite{gelenbe2013future}. As evacuees are commonly under pressure and likely to be confused, Ref. \cite{chittaro2008presenting} presents a location-aware 3D virtual building model to make use of the natural spatial abilities of mankind. The position of the evacuees is determined by a RFID system, while the 3D model is displayed by a 3D rendering engine. The system calculates the shortest path to the nearest exit and presents the path by using oriented arrows. Due to the low-accuracy of Wi-Fi positioning technology and high cost of a RFID system, Ref. \cite{ahn2011rescueme} proposes an infrastructure-less evacuation system based on image-based localization and augmented reality. Snapshots taken by evacuees are updated to cloud severs where an external image labeling service is used to identify landmarks. 3D evacuation tags are projected on the floor to guide evacuees to viable exits.

Augmented reality aided fire drills can reduce the high financial cost \cite{balasubramanian2006drillsim} and improve the training process quality. Ref. \cite{manasaugmented} uses a Wi-Fi position system to locate users and insert virtual fires into the live images of mobile phones. Ref. \cite{balasubramanian2006drillsim} presents a hybrid world that consists of a simulated world and a real world monitored by a smart space. Civilians that take part in a drill can immerse in the hybrid world through GUIs of mobile devices. However, current augmented-reality approaches do not consider the interaction between virtual objects and real world. For instance, the health value of a civilian should decrease when traversing through a virtual fire. Artificial agents should take appropriate intelligent behaviours when encountering a real person or other artificial entities. Ref. \cite{gelenbe2005simulating} proposes the idea to consider the interaction between virtual objects and real objects. A novel real-time moving-object injection method is presented based on image
segmentation and image registration techniques. Since behaviours of virtual object have a significant impact on the results of training personal or evaluating ``what if'' situation, three modules are designed to manage the movement of injected agents. The navigation module uses a RNN based algorithm to discover paths to destinations. The grouping module adopts the social potential fields approach \cite{reif1999social} to simulate the group behaviours. The imitation module allows agents to mimic the successful agents in missions. Based on this idea, we will investigate novel simulation models which combine with augmented reality to conduct simulation on realistic settings. Compared with traditional simulation models, this kind of models with real-life data tends to obtain more accurate results. Meanwhile, new augmented reality approaches to display virtual objects based on crowdsourced data \cite{stranders2011collabmap} will be studied to facilitate decision making for evacuees.

Another advantage of evaluating synthetic simulated conditions in physical world is to benefit current animal behaviour based crowd evacuation models \cite{zheng2009modeling}. The use of animals such as ants \cite{shiwakoti2010biologically} or mice \cite{saloma2003self} to simulate human being is due to the lack of real-life data during emergency and panic situations. However, the react of animals may differentiate from human-being. Therefore, employing augmented reality to crowed models can gain more precise results for crowd behaviours.

\vspace{0.5em}
\noindent\textbf{Improve security of network based emergency evacuation systems}
\vspace{0.5em}

As a low-cost and unobtrusive supplement of terrorist attacks, intervention on the emergency evacuation systems is a effective way to increase fatalities. Most of the current emergency evacuation systems are based on wireless sensor networks, which are susceptible to attacks such as node capture, radio frequency jamming, physical tampering and denial of service (DoS) \cite{perrig2004security}.

To the best of our knowledge, although a large amount of research has discussed the network security issues, few has investigated in the context of evacuation. Other than using the standard cryptography defense mechanism, Ref. \cite{gorbil2012resilience} incorporates identity-based signatures \cite{shamir1985identity} and content-based message verification to detect malicious attacks. The validity of message is determined by the information consistency and the fact that captured nodes create more malicious messages than non-malicious nodes. Three types of node misbehaviour that include drop-all, radio frequency jamming and false info are tested in a opportunistic communication based evacuation scenario and the results shows that the defense mechanism can effectively detect and resist malicious attacks.

Due to the limited computing power, memory, bandwidth and battery power, wireless sensor networks based emergency response systems are especially vulnerable to denial-of-service and denial-of-sleep attacks \cite{raymond2008denial}. Moreover, the development of cloud based emergency evacuation systems makes DoS an effective approach to limit the function of the systems by remote terrorists.

Ref. \cite{gelenbe2007self} proposes a mathematical model to simulate the behaviours of network flow at routers. A router is viewed as a single server queue and the probabilities of false alarm of a normal packet, correct detection of a malicious packet and buffer overflow are considered. Experimental results indicate that the model can reduce the effort of false alarms on non-malicious packets. To decrease the impact of false alarms more effectively, a attribute namely ``priority level'' is assigned to each packet based on its performance on validity test. Different from directly dropping, packets with low priority can receive service when there is available bandwidth.

The CPN protocol can naturally benefit the process of defending DoS or distributed DoS because SPs and ACKs store all intermediate nodes from source to destination. Hence, the potential malicious nodes can be located easily. All the nodes between a malicious source and the target node can drop attacking traffic to avoid useless occupancy of bandwidth. Meanwhile, as CPN is a QoS-driven protocol, packets can determine the validity by QoS and a node can allocate different bandwidth for flow of different QoS. In the future research, we will exploit novel mathematical models and mechanisms to improve network security in emergency evacuation. Additionally, new classification technologies to maximise detection accuracy of malicious packets based on Bayesian decision making will be investigated.

\vspace{0.5em}
\noindent\textbf{Apply G-networks model to evacuation process}
\vspace{0.5em}

One disadvantage of using M/M/1 model is that the network could only be considered as a network of queues other than a queueing network. Hence, interactions between neighbour nodes as well as the re-routing procedures of evacuees are not taken into account. G-networks \cite{gelenbe1994g} is a product form queueing networks inspired by \cite{gelenbe1989random}. It includes abundant elements such as positive customers, negative customers \cite{gelenbe1991product}, triggers \cite{gelenbe1993g}, resets \cite{gelenbe2002g} and batch removal \cite{gelenbe1993g}. As an effective tool for computer system modelling, it has been widely used in a large amount of applications such as network routing control \cite{morfopoulou2011network}, realising energy efficiency in packet networks \cite{morfopoulouimproving,gelenbe2011routing,gelenbe2011framework}, energy packet networks \cite{gelenbequality,gelenbe2012energy,gelenbe2012energystorage}, describing the workload in computer systems \cite{onvural1995asynchronous,gelenbe1998introduction} and modelling populations of agents and viruses \cite{gelenbe2007dealing}.

In the future research we will employ a G-networks model with positive customers and triggers to simulate a evacuation process. Positive customers represent evacuees and triggers depict re-routing decisions. The building can be divided into several Jackson networks where rooms are considered as sources and staircases and exits represent sinks. By applying the G-networks model, we could obtain interesting functions and parameters such as the average queue length of a node, the steady-state probability that a node has at least one evacuee, etc.

\vspace{0.5em}
\noindent\textbf{Investigate decision support algorithms in degraded conditions}
\vspace{0.5em}

Most of the previous emergency navigation algorithms do not consider malfunction of communication infrastructure and sensors which can be caused by the on-going hazard or malicious attacks. The communication infrastructure may be cut off by a hazard and cause a centralized algorithm invalid. Opportunistic communication and delay tolerant networks have been investigated to increase the resilience of emergency management systems \cite{bruno2008opportunistic,jiang2011adaptive,vahdat2000epidemic}. Ref. \cite{filippoupolitis2008emergency,filippoupolitis2011autonomous,gorbil2011opportunistic,gorbil2012intelligent,gelenbe2012wireless} has proposed a resilient emergency support system (ESS) with the aid of opportunistic communications \cite{pelusi2006opportunistic}. This system consists of SNs and CNs. SNs can detect the hazard in its vicinity and inform the location to the evacuees passing by. CNs are portable devices which are taken by occupants. The civilians will initially wander in the environment and can only inform the hazard to the civilians within its communication range. Each civilian calculates the shortest path to the exits with Dijkstra's algorithm. However, location tags may fail because of hazard and the portable devices may suffer from battery depletion as the power consumption for the flooding-based routing algorithm is relatively high. We believe that the CPN based algorithm can benefit this process by combining with opportunistic communication in at least two ways: first, other than epidemic routing algorithms, CPN unicasts most of time and could take the remaining battery power into account by using a specific goal function \cite{gelenbe2004power}; second, combining with the indoor localization technologies and a compound goal function, evacuees can coordinate their movements and form an ``evacuee network'' which can provide more sustainable connections with the survived communication infrastructures and realise coordinate localization for lost evacuees where sensor tags are destroyed. Each evacuees in the ``evacuee network'' can construct a ``cognitive map'' based on its own observation and information exchanges between civilians to make decision. The movement of evacuees can be determined by the grouping behaviours, imitation behaviours as well as its own decisions \cite{gelenbe2005simulating}.

\vspace{0.5em}
\noindent\textbf{Realise energy efficiency management of power distribution in emergency}
\vspace{0.5em}

Failure of power sources is common in an emergency situation and is obviously harmful to the evacuees. For instance, lighting failure could slow down the evacuees and even cause congestion. Interruption of power supply prohibits the usage of elevators and can decrease the efficiency of the evacuation procedure and aggravate the difficulties of evacuation for certain categories of evacuees (e.g. disabled people). Computers without battery may also have important information that needs to be saved and uploaded to the cloud before the fire reaches. Meanwhile, there are commonly some storage units (such as uninterruptible power supply (UPS)) in a modern building which can be used to sustain the electricity flow and support evacuation procedure when the electricity system is shorten. Currently research such as energy efficient buildings and smart buildings \cite{aguilar1997task,doukas2007intelligent,agarwal2010occupancy,lu2010smart,snoonian2003smart} has been focused on energy efficiency scheduling in standard conditions and few has concentrated on the energy distribution in an emergency evacuation. To the best of our knowledge, Only Ref. \cite{snoonian2003smart} presents a priority-scheduling algorithm for devices and sets a emergency signal to a higher priority to override the current schedule when a hazard happens. Meanwhile, with the development of smart building and energy-efficient building, the interoperation between different infrastructures and appliances tends to be possible. Inspired by robust infrastructure designing\cite{desmet2013interoperating} and energy packet networks \cite{gelenbequality,gelenbe2012energy,gelenbe2012energystorage}, an emergency electricity distribution and management system will be investigated to distribute electricity to diverse task efficiently in an emergency environment. G-networks model will be employed because it can predict the number of evacuee at a certain node and benefit the electricity distribution process. As Random neural networks has been proved to be a efficient tool in emergency task assignment\cite{gelenbe2010fast,gelenbe2010random,gelenbe2010randomemergency}, Random neural networks with multiple classes of signals\cite{gelenbe1999random,atalay1992parallel} will also be investigated because multiple classes of signals can represent different elements such as evacuees, rescuers, etc.

\renewcommand{\nomname}{List of Abbreviations}
\nomenclature{CPN}{Cognitive Packet Network}
\nomenclature{QoS}{Quality of Service}
\nomenclature{RNN}{Random Neural Network}
\nomenclature{SP}{Smart Packet}
\nomenclature{DP}{Dumb Packet}
\nomenclature{ACK}{Acknowledgement Packet}
\nomenclature{SN}{Sensor Node}
\nomenclature{DN}{Decision Node}
\nomenclature{CN}{Mobile Communication Node}
\nomenclature{DBES}{Distributed Building Evacuation Simulator}
\nomenclature{DoS}{Denial of Service}
\printnomenclature[2.5cm]

\bibliography{esa}
\end{document}